\documentclass[aps,reprint,superscriptaddress,nofootinbib,longbibliography,notitlepage,floatfix]{revtex4-2}
\usepackage{float}

\usepackage{graphicx}
\usepackage{tabularx}
\usepackage{amsmath}
\usepackage{amstext}
\usepackage{amssymb}
\usepackage{xfrac}
\usepackage[colorlinks,citecolor=blue]{hyperref}
\usepackage{graphicx}
\usepackage{amsmath}
\usepackage{amstext}
\usepackage{amssymb}
\usepackage{amsfonts}
\usepackage{longtable,booktabs}
\usepackage{hyperref}
\usepackage{url}
\usepackage{subfigure}
\usepackage{dsfont}
\usepackage{booktabs}
\usepackage{amsbsy}
\usepackage{dcolumn}
\usepackage{amsthm}
\usepackage{bm}
\usepackage{esint}
\usepackage{multirow}
\usepackage{hyperref}
\usepackage{cleveref}
\usepackage{amsmath}
\usepackage{amsfonts}
\usepackage{amssymb}
\usepackage{mathrsfs}
\usepackage{amsfonts}
\usepackage{amsbsy}
\usepackage{dcolumn}
\usepackage{bm}
\usepackage{multirow}
\usepackage{color}
\usepackage{extarrows}
\usepackage{datetime}
\usepackage{comment}
\usepackage[super]{nth}
\usepackage{tikz-cd}

\usepackage{scalerel}
\usepackage{tikz}
\usetikzlibrary{svg.path}

\usepackage{algorithm}
\usepackage{algpseudocode}

\usepackage{xr}

\definecolor{orcidlogocol}{HTML}{A6CE39}
\tikzset{
	orcidlogo/.pic={
		\fill[orcidlogocol] svg{M256,128c0,70.7-57.3,128-128,128C57.3,256,0,198.7,0,128C0,57.3,57.3,0,128,0C198.7,0,256,57.3,256,128z};
		\fill[white] svg{M86.3,186.2H70.9V79.1h15.4v48.4V186.2z}
		svg{M108.9,79.1h41.6c39.6,0,57,28.3,57,53.6c0,27.5-21.5,53.6-56.8,53.6h-41.8V79.1z M124.3,172.4h24.5c34.9,0,42.9-26.5,42.9-39.7c0-21.5-13.7-39.7-43.7-39.7h-23.7V172.4z}
		svg{M88.7,56.8c0,5.5-4.5,10.1-10.1,10.1c-5.6,0-10.1-4.6-10.1-10.1c0-5.6,4.5-10.1,10.1-10.1C84.2,46.7,88.7,51.3,88.7,56.8z};
	}
}

\newcommand\orcidicon[1]{\href{https://orcid.org/#1}{\mbox{\scalerel*{
				\begin{tikzpicture}[yscale=-1,transform shape]
					\pic{orcidlogo};
				\end{tikzpicture}
			}{|}}}}

\hypersetup{
    colorlinks=magenta,
    linkcolor=blue,
    filecolor=magenta,
        urlcolor=magenta,
}

\newcommand{\im}{\text{im}}

\def\C{\mathcal{C}}

\def\K{\mathcal{K}}

\usepackage{extarrows}

\usepackage{datetime}

\usepackage{comment}

\usepackage{lineno}

\begin{document}
	
	
	\title{Quantum bootstrap product codes}
	
	
	
	\author{Meng-Yuan Li\orcidicon{0000-0001-8418-6372}}
	\affiliation{Institute for Advanced Study, Tsinghua University, Beijing, 100084, China}
	\email{my-li@mail.tsinghua.edu.cn}

 	\date{\today}
\begin{abstract}
	
	Product constructions constitute a powerful method for generating quantum CSS codes, yielding celebrated examples such as toric codes and asymptotically good low-density parity check (LDPC) codes. Since a CSS code is fully described by a chain complex, existing product formalisms are predominantly homological, defined via the tensor product of the underlying chain complexes of input codes, thereby establishing a natural connection between quantum codes and topology. In this Letter, we introduce the \textit{quantum bootstrap product} (QBP), an approach that extends beyond this standard homological paradigm. Specifically, a QBP code is determined by solving a consistency condition termed the ``bootstrap equation''. We find that the QBP paradigm unifies a wide range of important codes, including general hypergraph product (HGP) codes of arbitrary dimensions and fracton codes typically represented by the X-cube code. Crucially, the solutions to the bootstrap equation yield chain complexes where the chain groups and associated boundary maps consist of multiple components. We term such structures \textit{fork complexes}. This structure elucidates the underlying topological structures of fracton codes, akin to foliated fracton order theories. Beyond conceptual insights, we demonstrate that the QBP paradigm can generate self-correcting quantum codes from input codes with constant energy barriers and surpass the code-rate upper bounds inherent to HGP codes. Our work thus substantially extends the scope of quantum product codes and provides a versatile framework for designing fault-tolerant quantum memories.
	
\end{abstract}


\maketitle

\textit{Introduction.}
Based on extensive exploration in both theoretical studies and experimental realizations, quantum error correction (QEC) serves as a promising approach for realizing large scale practical quantum computation~\cite{Steane1996,Shor1996,Calderbank1996,Gottesman1997,Dennis2002,Kitaev2003,Beugnon2007,RyanAnderson2021,Krinner2022,GQA2023,Sivak2023,Bluvstein2024,Bravyi2024,Gupta2024,Paetznick2024,Reichardt2024,GQA2025,Putterman2025,SalesRodriguez2025}. As an important QEC schemes, stabilizer codes have attracted significant attention across diverse fields of physics beyond quantum information, including condensed matter, high energy physics, and mathematical physics~\cite{Steane1996,Shor1996,Calderbank1996,Gottesman1997,Dennis2002,Kitaev2003,Wen2003a,Fattal2004,Kitaev2006a,Levin2006,Bullock2007,Haah2013,Vijay2015,Vijay2016,Pretko2018,Li2020,Liu2023,Li2025}. Consequently, systematic methods to construct stabilizer codes are highly desired for both theoretical understanding and practical realization of QEC. In this context, quantum product codes have emerged as a powerful paradigm. Based on the chain complex representations of classical and quantum Calderbank–Shor–Steane(CSS)-type stabilizer codes, a natural strategy is to take the tensor product of chain complexes representing classical codes to obtain chain complexes representing CSS codes, leading to the seminal hypergraph product (HGP) formalism~\cite{Tillich2009}. Since the advent of HGP codes, a series of quantum product code formalisms have been proposed, such as the quantum lifted product, fiber bundle product, and balanced product, among others~\cite{Kovalev2012,Bravyi2013,Fan2016,Zeng2019a,Hastings2020,Panteleev2020,Breuckmann2020,Leverrier2022,Ostrev2024,Rakovszky2024,Tan2025}. Numerous important quantum codes, including toric codes and asymptotically good low-density parity check (LDPC) codes, have been obtained as product codes, demonstrating the fascinating utility of product code constructions.

In this Letter, we propose a formalism that extends beyond the chain complex product paradigm, termed the \textit{quantum bootstrap product} (QBP). We focus on quantum CSS codes, which are mathematically described by length-$3$ chain complexes $\C: C_2\xrightarrow{\partial_2} C_1\xrightarrow{\partial_1} C_0$ satisfying the condition $\partial_1 \partial_2 = 0$~\cite{Gottesman1997,Kitaev2003,Bombin2007a}. In the QBP paradigm, we employ the tensor product of input classical codes represented by length-$2$ complexes $\C^i: C_1^i\xrightarrow{\delta^i} C^i_0$ to construct only a length-$2$ segment $C_1\xrightarrow{\partial_1} C_0$ of the target product code. Subsequently, we solve a \textit{bootstrap equation} derived from the chain complex condition $\partial_1 \partial_2 = 0$ to find the valid boundary operator $\partial_2$, thereby completing the complex $C_2\xrightarrow{\partial_2} C_1\xrightarrow{\partial_1} C_0$, justifying the name. The QBP naturally reproduces the HGP formalism and its high-dimensional generalizations, including the celebrated self-correcting 4D toric code~\cite{Dennis2002,Tillich2009, Zeng2019a}, while offering a more versatile framework.

Crucially, the QBP circumvents the code-rate limitations of HGP codes. For instance, we can consider classical 1D repetition codes of parameters $[n=L,k=1,d=L]$, denoted by $\{\mathcal{C}^{1D}_i\}$, as input codes, where $i$ is an index. Here we adopt the standard representation of code parameters, using $[[n,k,d]]$ and $[n,k,d]$ for quantum and classical codes, respectively. $n$, $k$, and $d$ refer to the number of physical qubits, the number of logical qubits (code dimension), and code distance, respectively. The high-dimensional HGP of $\{\mathcal{C}^{1D}_i\}$ can generate toric codes in arbitrary dimensions, where the code dimension $k$ is constrained to be a constant~\cite{Tillich2009,Zeng2019a}. Meanwhile, the QBP of $\{\mathcal{C}^{1D}_i\}$ extends to the realm of \textit{fracton codes}, where both the code dimension and code distance grow polynomially with the number of physical qubits~\cite{Vijay2015,Vijay2016,Ma2018a,Tian2019,Yoshida2013,Chamon2005,Haah2011,Song2022}. Such fracton codes are characterized by more complicated syndrome distributions than pure topological codes, which potentially lead to better error thresholds~\cite{Song2022,Canossa2025}. We concretely demonstrate this by constructing \textit{tetra-digit codes}, a systematically constructed family of codes including the distinguished X-cube fracton code, as QBP codes from $\{\mathcal{C}^{1D}_i\}$~\cite{Li2020,Li2021,Li2023,Hu2025}. In particular, the code dimensions of a subset of tetra-digit codes are proved to be $\Theta(n^{1-2/p})$, explicitly surpassing the HGP codes from $\{\mathcal{C}^{1D}_i\}$~\cite{Li2021} ($p$ is the number of input codes). Consequently, the QBP paradigm substantially expands the scope of quantum product codes.

\begin{figure}[t!]
	\centering
	\includegraphics[width=1.0\columnwidth]{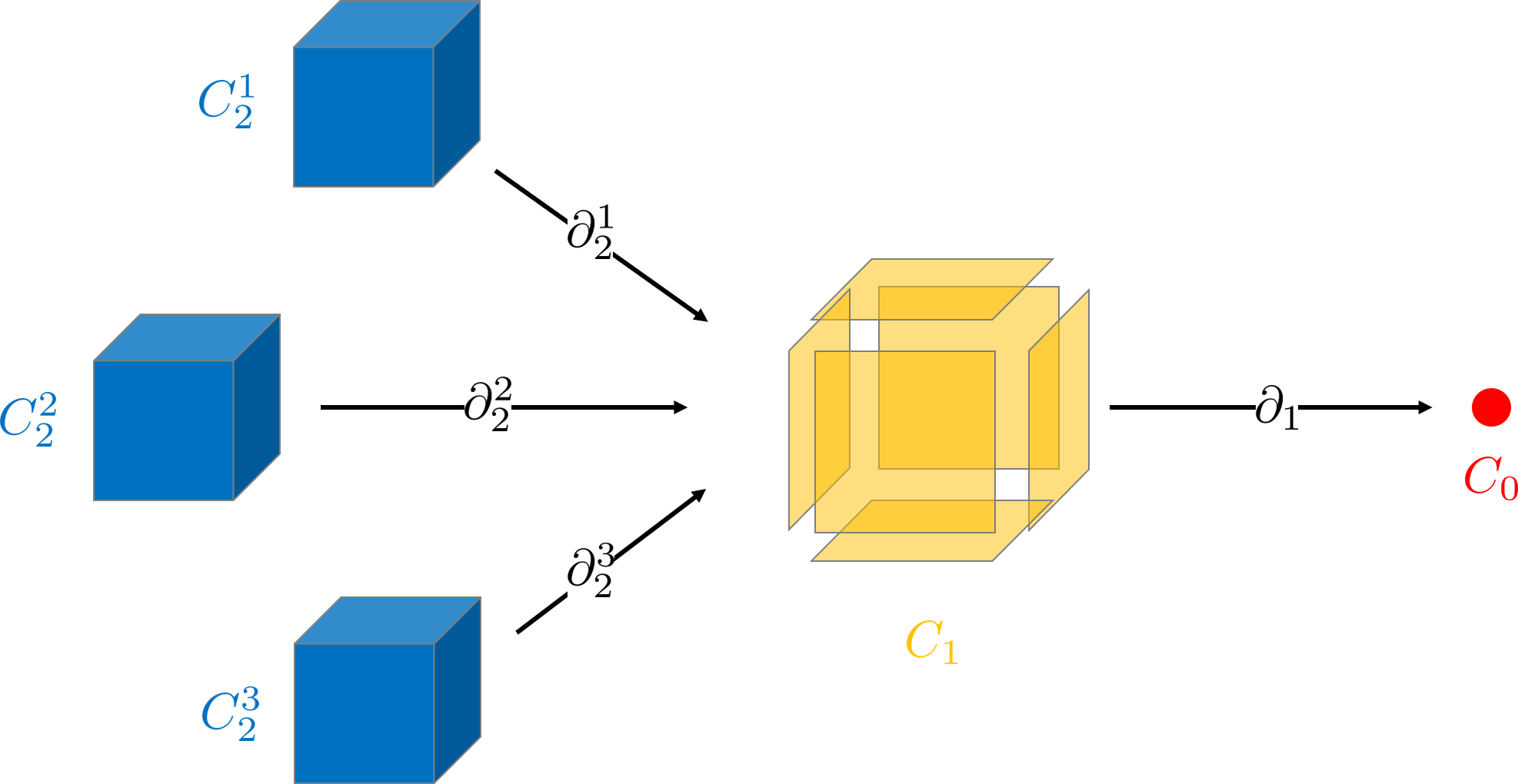}
	\caption{A schematic picture of fork complexes. For clarity, here we draw a fork complex based on the chain complex representing a 3D cubic lattice. The 1-chain group $C_1$ and 0-chain group $C_0$ respectively correspond to 2-cells (plaquettes) and 0-cells (vertices) of the lattice, while the 2-chain group $C_2$ is the direct sum of three copies of the 3-cells (cubes); each copy $C^i_2$ is equipped with a unique boundary operator $\partial^i_2$ mapping to $C_1$.}
	\label{fig:fork}
\end{figure} 

Beyond code construction, the QBP paradigm also offers insight into quantum many body theories. Generally, the chain complex serves as a bridge between topological CSS codes and topological orders~\cite{Kitaev2003,Wen2003a,Bombin2007a}. However, as an exotic kind of topological orders, fracton orders realized in fracton codes blur this correspondence~\cite{Chamon2005,Haah2011,Vijay2015}. While many fracton codes are defined on manifolds, their chain complex representations often lack a manifest connection to the underlying topology. The QBP shed light on this ambiguity: by solving the bootstrap equation, we find that the boundary operator $\partial_2$ typically decomposes into multiple components $\partial_2^l$ (where $l$ indexes the component), each satisfying $\partial_1 \partial^l_2 = 0$ for a subgroup $C^l_2 \subset C_2$, where the total $C_2$ is the direct sum of all $C^l_2$ (see Fig.~\ref{fig:fork} for a schematic representation). We term such a structure a \textit{fork chain complex}, or \textit{fork complex} for brevity, when it cannot be described by a segment of the tensor product chain complex of input codes. Each branch $C^l_2\xrightarrow{\partial^l_2} C_1 \xrightarrow{\partial_1} C_0$ constitutes a valid chain complex in itself. We also use $(C^l_2\xrightarrow{\partial^l_2})^{\bigoplus_l} C_1 \xrightarrow{\partial_1} C_0$ to represent a chain complex to emphasize its fork structure. Such a fork structure reveals the underlying topological dependence of fracton codes, akin to the foliated fracton order theory~\cite{Shirley2018,Shirley2019,Shirley2019a,Slagle2021}.

\textit{Notations and conventions.}
Throughout this Letter, we mainly utilize the products of classical 1D repetition codes with $L$ bits as concrete examples, thus it is convenient to introduce some notations. Such a 1D repetition code is denoted as $\C^{1D}$. Defined on a 1D lattice of $L$ bits with periodic boundary conditions (PBC), $\C^{1D}$ is characterized by the parameters $[n=L, k=1, d=L]$. Its chain complex representation is given by a length-$2$ chain complex $C_1 \xrightarrow{\delta} C_0$, where $C_1$ and $C_0$ are vector spaces over $\mathbb{F}_2$ spanned by bits (on links) and checks (on vertices), respectively. Note that as we focus on chain complexes representing codes in this work, all chain groups are assumed to be vector spaces over $\mathbb{F}_2$ by default.

We frequently consider the tensor product of classical codes, denoted as $\C^{TP}$. When input codes are $p$ copies of $\C^{1D}$, this product also represents a $p$-dimensional hypercubic lattice of size $L^p$, denoted by $\K$. The equivalence is established by identifying the chain groups of $\C^{TP}$ with the cells of $\K$: the $k$-th chain group $C^{TP}_k$ is spanned by the $k$-cells of the lattice (e.g., vertices for $C^{TP}_0$, links for $C^{TP}_1$, plaquettes for $C^{TP}_2$, etc.). The boundary operator of the tensor product, $\partial^{TP}_k = \sum_{i=1}^p \delta^i$ (where $\delta^i$ acts on the $i$-th input code), exactly reproduces the homological boundary map $\partial^{\K}_k$, which maps a $k$-cell to the sum of its boundary $(k-1)$-cells. By formally setting $\delta^i c = 0$ for an arbitrary check $c\in C^i_0$, such polynomial representations faithfully reproduce the boundary operators. See Supplementary Materials (SM) for more details~\cite{supp}.

\textit{Quantum bootstrap.}
A QBP formalism is specified by a triple of integers $(p,q,w)$ with $p>q>w$. For convenience, we use $\mathcal{C}$ to denote both a code and its corresponding chain complex where no ambiguity arises. A $(p,q,w)$ QBP code is constructed from $p$ classical linear codes $\mathcal{C}^i:C^i_1 \xrightarrow{\delta^i} C^i_0$, indexed by $I_p = \{1,2,\cdots,p\}$. The resulting product code is described by a complex:
\begin{align}
	\mathcal{C}^{(p,q,w)}:C^{(p,q,w)}_2 \xrightarrow{\partial^{(p,q,w)}_2} C^{(p,q,w)}_1 \xrightarrow{\partial^{(p,q,w)}_1} C^{(p,q,w)}_0.
\end{align} 
The construction proceeds in two steps.

First, we define the qubits and $X$-checks using the tensor product of the input codes, $\mathcal{C}^{TP}$. We designate the $q$-chains and $w$-chains of $\mathcal{C}^{TP}$ to constitute the qubit space ($C^{(p,q,w)}_1$) and $X$-check space ($C^{(p,q,w)}_0$), respectively:
\begin{align}
	C^{(p,q,w)}_1 &= C^{TP}_q =  \bigoplus_{S\subset I_p, |S|=q} (\bigotimes_{i\in S}C^i_1)\otimes (\bigotimes_{j\in I_p\setminus S}C^j_0), \\
	C^{(p,q,w)}_0 &= C^{TP}_w = \bigoplus_{S\subset I_p, |S|=w} (\bigotimes_{i\in S}C^i_1)\otimes (\bigotimes_{j\in I_p\setminus S}C^j_0).
\end{align}
The boundary map $\partial^{(p,q,w)}_1: C^{(p,q,w)}_1 \to C^{(p,q,w)}_0$ is defined as the sum of all products of $\delta^i$ lowering the degree from $q$ to $w$. In the polynomial representation, where $\delta^i$ denotes the boundary operator of the $i$-th input code $\C^i$, this map corresponds to the homogeneous polynomial:
\begin{align}
	\partial^{(p,q,w)}_1 = \sum_{S\subset I_p, |S|=q-w} \prod_{i\in S} \delta^i.
\end{align} 
This polynomial represents the sum of tensor products of boundary operators. For instance, the boundary maps $\partial^{TP}_k$ of the tensor product complex are all represented by the polynomial $\sum_{i=1}^p \delta^i$, regardless of the degree $k$.

Second, unlike $\partial^{(p,q,w)}_1$, the boundary operator $\partial^{(p,q,w)}_2$ is not directly obtained from the tensor product structure. Instead, it is determined by solving a consistency condition termed the \textit{bootstrap equation}. We seek solutions $\partial^l_2$ (where $l$ indexes the solution) that map from a space $C^{TP}_t$ of degree $t$ (with $t>q$) to the qubit space $C^{(p,q,w)}_1$. The bootstrap equation enforces the chain complex condition (ensuring the commutativity of checks):
\begin{align}
	R_{I^l_t}[\partial^{(p,q,w)}_1] \partial^{l}_2 = 0.
\end{align}
Here, $I^l_t\subset I_p$ is a specific subset of indices with size $t$, and $R_{I^l_t}[\partial^{(p,q,w)}_1]$ denotes the \textit{restriction} of $\partial^{(p,q,w)}_1$ to $I^l_t$. The restriction is defined as the projection of the polynomial $\partial^{(p,q,w)}_1$ onto the subalgebra generated by $\{\delta^i \mid i\in I^l_t\}$; this operation isolates terms in $\partial^{(p,q,w)}_1$ composed exclusively of $\delta^i$ with indices in $I^l_t$. A valid solution $\partial^{l}_2$ is a homogeneous polynomial of degree $t-q$ supported exclusively on $I^l_t$. A general algorithm to solve bootstrap equations is provided in SM~\cite{supp}.

Each solution $\partial^{l}_2$ specifies a distinct subgroup $C^l_2 \subset C^{TP}_t$ supported on $I^l_t$:
\begin{align}
	C^l_2 = (\bigotimes_{i\in I^l_t}C^i_1)\otimes (\bigotimes_{j\in I_p\setminus I^l_t}C^j_0),
\end{align} 
satisfying $\partial^{(p,q,w)}_1 \partial^l_2 c_2 = 0$ for all $c_2 \in C^l_2$. The total $Z$-check space is the direct sum $C^{(p,q,w)}_2 = \bigoplus_l C^l_2$. We term a chain complex with multiple $\partial^l_2$ supported on identical index sets $I^l_t$ a \textit{fork complex}. The resulting code space corresponds to the homology group $H^{(p,q,w)}_1=\ker{\partial^{(p,q,w)}_1} / \sum_{l} \im \partial^l_2$.

To maximize the code distance, one should include all independent solutions to the bootstrap equation. Note that solutions may be linearly dependent. For example, if $\partial^1_2 = \delta^1 + \delta^2$, $\partial^2_2 = \delta^1 + \delta^3$, and $\partial^3_2 = \delta^2 + \delta^3$ are solutions for the same index set $I^1_3 = I^2_3 = I^3_3 = \{1,2,3\}$, the relation $\partial^1_2 + \partial^2_2 = \partial^3_2$ renders $\partial^3_2$ redundant, which means that the corresponding subgroup $C^3_2$ is spanned by redundant $Z$-checks. In this work, we consider sets of independent solutions by default.

\textit{Self-correcting QBP codes.}
The QBP framework naturally includes HGP codes of arbitrary dimensions~\cite{Tillich2009,Zeng2019a}. More concretely, a $(p,q,w)$ QBP reduces to a $p$-dimensional HGP when $q-w=1$. Consequently, the QBP can also generate self-correcting codes from classical input codes with constant energy barriers, which distinguishes it from conventional two-dimensional HGP constructions~\cite{Zhao2025,Zhao2025a}. As a concrete example, we generate the 4D toric code as a $(4,2,1)$ QBP code of four $\C^{1D}$.

First, we consider the $(4,2,1)$ QBP of four general classical linear codes $\mathcal{C}^i$ ($i\in I_4=\{1,2,3,4\}$). Following the general formalism, we identify the qubits and $X$-checks by $C^{(4,2,1)}_1 = C^{TP}_2$ and $C^{(4,2,1)}_0 = C^{TP}_1$. The boundary map is given by the polynomial $\partial^{(4,2,1)}_1 = \sum_{i=1}^4 \delta^i$. By solving the bootstrap equation, we find that for each subset $I^l_3 \subset I_4$ of size three, there exists only one solution $\partial^l_2 = \sum_{j \in I^l_3} \delta^j$~\cite{supp}. The corresponding $Z$-check spaces are $C^l_2 = (\bigotimes_{i\in I^l_3} C^i_1) \otimes (\bigotimes_{j\in I_4\setminus I^l_3} C^j_0)$. 

\begin{figure}[t!]
	\centering
	\includegraphics[width=1.0\columnwidth]{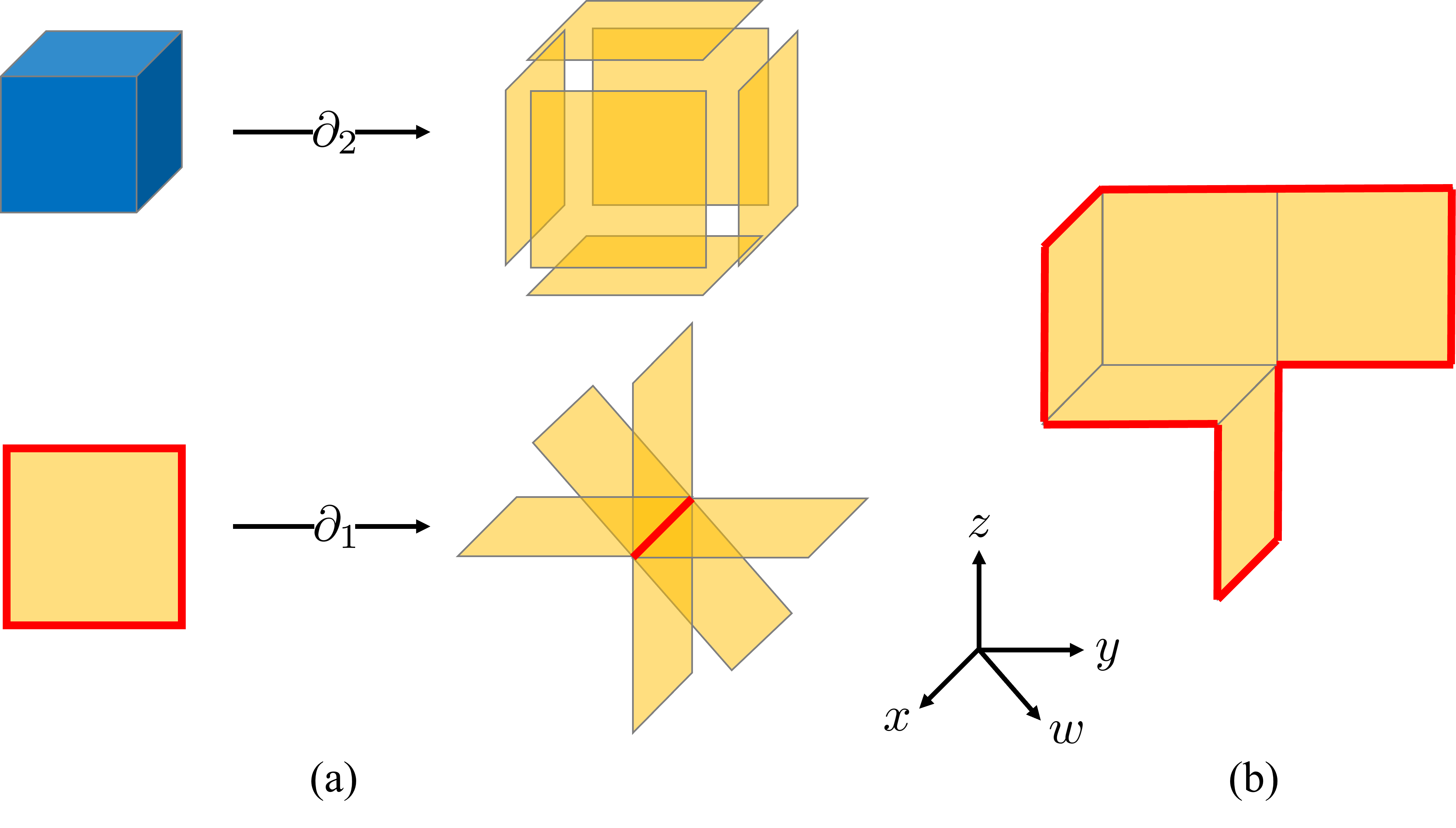}
	\caption{Chain complex representation of the 4D toric code. In (a), we demonstrate the boundary operators $\partial_2$ and $\partial_1$ defining $Z$-checks and $X$-checks, respectively. More concretely, $\partial_2$ maps a $Z$-check on a cube to the six qubits on the plaquettes surrounding the cube, and $\partial_1$ maps a qubit on a plaquette to the four $X$-checks on the links surrounding the plaquette. Thus, an $X$-check on a link involves six qubits on plaquettes adjacent to the link. In (b), we demonstrate a loop-like syndrome (colored red) created by Pauli $Z$ errors distributed on a membrane (colored yellow). Such loop-like syndromes underpin the self-correcting property of the 4D toric code.}
	\label{fig:4dtc}
\end{figure} 

When the input codes are four identical $\mathcal{C}^{1D}$, the tensor product complex $\mathcal{C}^{TP}$ describes the cellulation of a four-dimensional torus $T^4$. The identifications $C^{(4,2,1)}_1 = C^{TP}_2$ and $C^{(4,2,1)}_0 = C^{TP}_1$ place qubits on the plaquettes ($2$-cells) and $X$-checks on the links ($1$-cells) of the 4D hypercubic lattice. The decomposition of the $Z$-check space $C^{(4,2,1)}_2 = \bigoplus_l C^l_2$ ensures that every cube ($3$-cell) hosts a $Z$-check, with the index set $I^l_3$ specifying the orientation of the cubes. Since both $\partial^{(4,2,1)}_1$ and $\partial^{(4,2,1)}_2$ are homogeneous polynomials of degree one, the resulting checks involve only nearest qubits. This construction exactly reproduces the 4D toric code, a paradigmatic example of a self-correcting quantum memory with thermal stability~\cite{Dennis2002,Alicki2008}, as illustrated in Fig.~\ref{fig:4dtc}.

Finally, we demonstrate that the QBP framework generally encompasses the HGP formalism of arbitrary dimensions. The HGP code of $p$ classical input codes $\C^i$ ($i=1,2,\cdots, p$) corresponds to a length-$3$ segment of their full tensor product complex $\mathcal{C}^{TP}$. Within the QBP formalism, setting $C^{(p,q,q-1)}_1 = C^{TP}_q$ and $C^{(p,q,q-1)}_0 = C^{TP}_{q-1}$ implies $\partial^{(p,q,q-1)}_1 = \partial^{TP}_q$. The bootstrap equation $R_{I^l_t}[\partial^{(p,q,q-1)}_1] \partial^l_2 = 0$ then admits a solution $\partial^l_2 = \sum_{j\in I_{q+1}} \delta^j$ for each subset $I_{q+1} \subset I_p$ of size $q+1$. Consequently, we recover $\partial_2 = \partial^{TP}_{q+1}$ and $C^{(p,q,q-1)}_2 = C^{TP}_{q+1}$, confirming that HGP codes are equivalent to $(p,q,w)$ QBP codes satisfying $q-w = 1$.

\textit{Fracton QBP codes.}
Furthermore, we demonstrate that the QBP paradigm can generate fracton codes. As a kind of topological codes, fracton codes also possess logical operators supported on topologically nontrivial objects, while they are characterized by code dimensions scaling with the system size $n$ and more complex syndrome distributions than pure topological codes, which potentially leads to improved error thresholds~\cite{Chamon2005,Haah2011,Vijay2015,Li2020,Song2022,Canossa2025}. For instance, we show that with $\C^{1D}$ as inputs, the $(p,q,0)$ QBP formalism generates the tetra-digit (TD) codes, a broad family of fracton codes in arbitrary dimensions labeled by four integers that specify the assignment of qubits and checks (hence the name; see SM~\cite{supp} for a brief review)~\cite{Li2020,Li2021,Hu2025}. This family includes the celebrated X-cube code~\cite{Vijay2016} and fracton codes with spatially extended (e.g., string-like) syndromes, and can formally include toric codes of all dimensions, making it a versatile platform to study both CSS codes and topological orders. For example, the 2D toric code, 3D toric code, and X-cube code are labeled by $[0,1,2,2]$, $[1,2,3,3]$, and $[0,1,2,3]$, respectively. Notably, while the code dimensions of HGP codes based on $\mathcal{C}^{1D}$ are limited to $k=\Theta(1)$, such TD codes can achieve a polynomial scaling of $k$, demonstrating that the QBP framework overcomes the code-rate limitations of the HGP construction.

First, we consider the $(p,q,0)$ QBP code $\mathcal{C}^{(p,q,0)}$ based on arbitrary classical input codes. Following the general formalism, the $X$-checks and qubits are identified with $C^{(p,q,0)}_0 = C^{TP}_0$ and $C^{(p,q,0)}_1 = C^{TP}_q$, respectively, with the boundary map $\partial^{(p,q,0)}_1 = \sum_{I_{q}} \prod_{i\in I_{q}} \delta^i$. By solving the bootstrap equation, we find that the solutions are generated by operators of the form $\partial^l_2 = \delta^i + \delta^j$, where $i,j$ are distinct indices from a subset $I_{q+1}\subset I_p$. For a given support $I_{q+1}$, there are $q$ independent solutions. This multiplicity of boundary operators is the hallmark of the fork complex structure.

When the input codes are $p$ identical $\mathcal{C}^{1D}$ with parameters $[L,1,L]$, where each is represented by a chain complex $C^i_1 \xrightarrow{\delta^i} C^i_0$, the resulting $(p,q,0)$ QBP code corresponds to the TD fracton code labeled by $[p-q-1,p-q,p-q+1,p]$. As an example, we first demonstrate that the X-cube code can be obtained by $(3,2,0)$ QBP. 

\begin{figure}[t!]
	\centering
	\includegraphics[width=1.0\columnwidth]{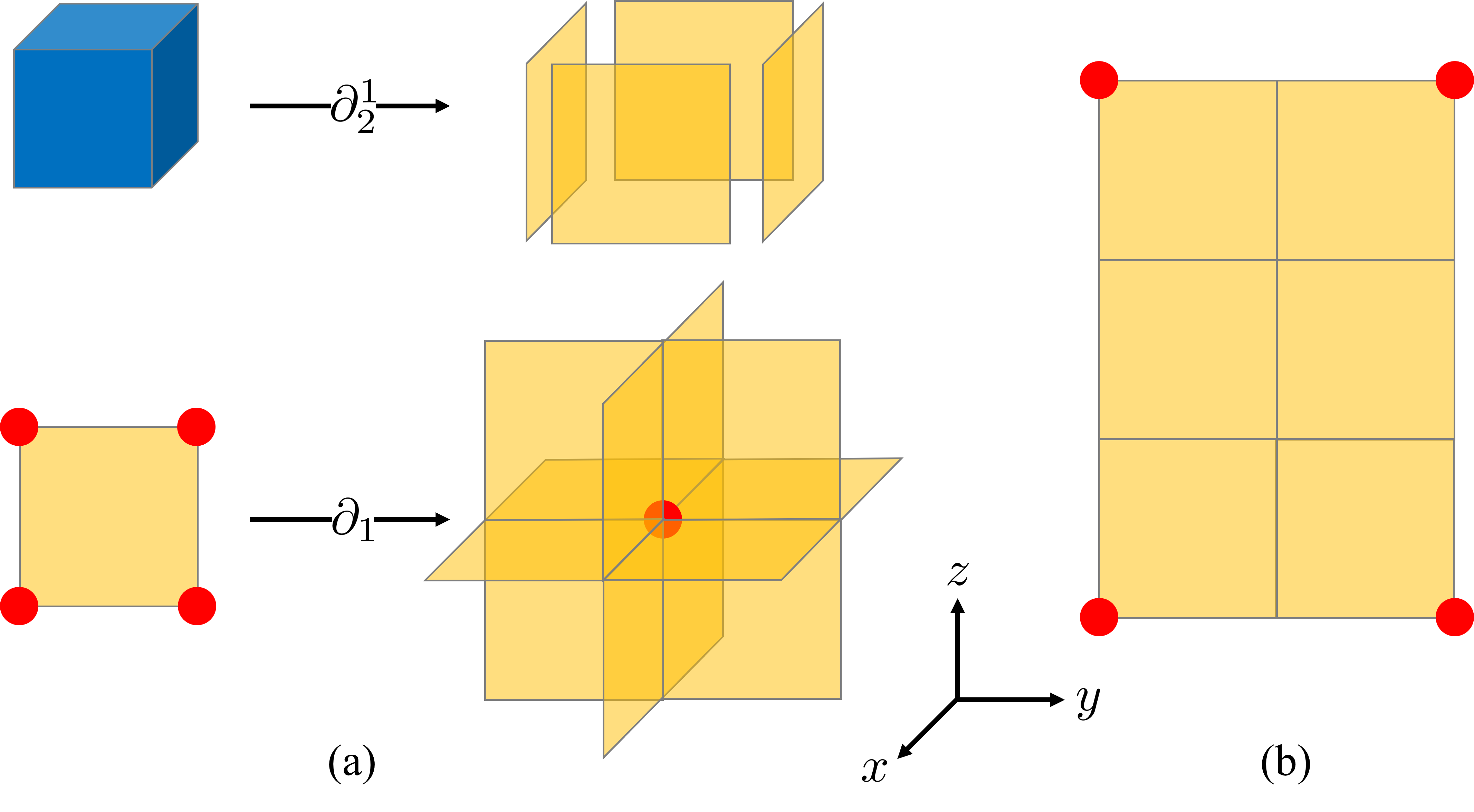}
	\caption{Fork complex representation of the X-cube code on the dual lattice. In (a), we illustrate the boundary operators $\partial_1$ and $\partial^1_2$ defining $X$-checks and one type of $Z$-checks, respectively ($\partial^2_2$ can be represented similarly). Specifically, $\partial^1_2$ maps a $Z$-check on a cube to four qubits on the plaquettes that are both adjacent to the cube and perpendicular to the $x$ or $y$ axis. Meanwhile, $\partial_1$ maps a qubit on a plaquette to the four $X$-checks on the vertices surrounding the plaquette. Thus, an $X$-check on a vertex involves twelve qubits on the nearest plaquettes. These checks exactly define the X-cube code on the dual lattice. In (b), we show four point-like syndromes (colored red) created by Pauli $Z$ errors distributed on a membrane (colored yellow). Such syndromes generated at the corners of membranes provide more redundant information for decoding than typical point-like syndromes generated at the endpoints of error strings.}
	\label{fig:xcube}
\end{figure} 

We denote the fork complex and standard chain complex representations of the $(3,2,0)$ QBP of three $\mathcal{C}^{1D}$ by $(C^l_2\xrightarrow{\partial^l_2})^{\bigoplus_l} C_1 \xrightarrow{\partial_1} C_0$ and $C_2 \xrightarrow{\partial_2} C_1 \xrightarrow{\partial_2} C_0$, respectively. According to the general formalism, we have $C_1 = C^{TP}_2$ and $C_0 = C^{TP}_0$. As $\C^{TP}$ represents a 3D cubic lattice with PBC, $C^{TP}_0$ and $C^{TP}_2$ are spanned by the vertices and plaquettes of the lattice, respectively, and the boundary operator relating them is $\partial_1 = \delta^1 \delta^2 + \delta^1 \delta^3 +\delta^2 \delta^3$. Consequently, the $X$-checks and qubits of the QBP code are respectively assigned to vertices and plaquettes. Specifically, for a given vertex, the corresponding $X$-check involves all twelve qubits on the plaquettes nearest to the vertex. The two independent solutions of the bootstrap equation can be chosen as $\partial^1_2 = \delta^1 + \delta^2$ and $\partial^2_2 = \delta^1 + \delta^3$, both are supported on the same index set $I_3$. Therefore, their associated subgroups of $C_2$ are both equivalent to $C^{TP}_3$. The total $Z$-check space is thus $C_2 = C^1_2 \oplus C^2_2 = C^{TP}_3 \oplus C^{TP}_3$. Namely, each cube is assigned two independent $Z$-checks; each check involves four plaquettes that are both nearest to the cube and centered on the same plane. This $(3,2,0)$ QBP code exactly reproduces the X-cube code on the dual lattice, whose original definition places qubits on links~\cite{Vijay2016, supp} (see Fig.~\ref{fig:xcube}).

\begin{figure}[t!]
	\centering
	\includegraphics[width=1.0\columnwidth]{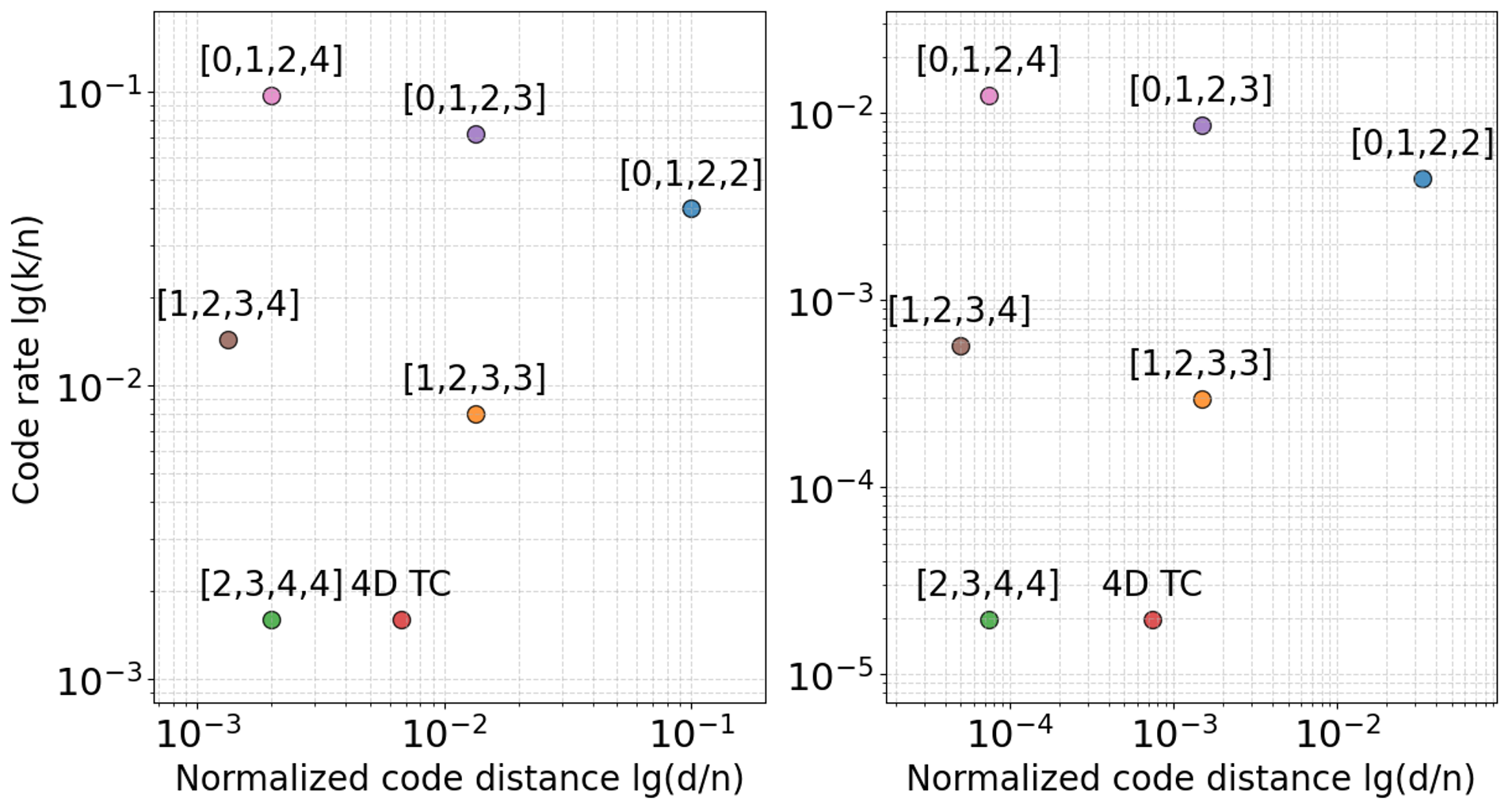}
	\caption{Comparison of code parameters for a series of QBP codes. All codes are defined on (hyper)cubic lattices of linear size $L$ with dimensions specified by the code; the left and right panels show $L=5$ and $L=15$, respectively. For clarity, we normalize $k$ and $d$ by considering $k/n$ and $d/n$ and plot the data on a logarithmic scale. As observed, while the code dimensions of $[p-2,p-1,p,p]$ codes ($p$-dimensional toric codes) obtained by the HGP of $\C^{1D}$ can only be constants, QBP-generated $[0,1,2,p]$ codes show superior performance in terms of code rates. Here, 4D TC refers to the 4D toric code with only loop-like syndromes, while $[2,3,4,4]$ refers to the 4D toric code with both point- and membrane-like syndromes.}
	\label{fig:comparision}
\end{figure}

We can now promote this geometric understanding to the general $(p,q,0)$ QBP code of $\C^{1D}$. Let $\K$ be the $p$-dimensional hypercubic lattice. This QBP code places $X$-checks on vertices ($0$-cells, $C_0=C^{TP}_0$) and qubits on $q$-cells ($C_1=C^{TP}_q$) of $\K$. An $X$-check involves all qubits on the $q$-cells incident to the vertex. For the $Z$-checks, the independent solutions $\delta^{i} + \delta^{j}$ correspond to checks residing on $(q+1)$-cells. Specifically, each $(q+1)$-cell in $\K$ is assigned $q$ independent $Z$-checks, where each check acts on four qubits that are both incident to the $(q+1)$-cell and centered on the same $(q+1)$-dimensional hyperplane. This construction exactly reproduces the TD code labeled $[p-q-1,p-q,p-q+1,p]$ on the dual lattice~\cite{Li2020,supp}. Crucially, when $p-q=1$, the code parameters scale as $[[pL^p,\Theta(L^{p-2}),L]]$~\cite{Li2021}. This polynomial scaling of the code dimension $k$ confirms that QBP codes can overcome the code rate limitation of the HGP framework (where $k=\Theta(1)$ for $\C^{1D}$ inputs), as illustrated in Fig.~\ref{fig:comparision}.

Finally, we demonstrate that the QBP paradigm reveals topological structures within fracton TD codes. Specifically, a chain complex representing a toric code emerges by solving the bootstrap equation. Without loss of generality, consider $I^1_{q+1} = \{1,2,\cdots,q+1\}$ and the solution $\partial^{1}_2 = \delta^1 + \delta^2$. In this case, the restricted boundary map is $R_{I^1_{q+1}}[\partial^{(p,q,0)}_1] = \sum_{I_{q} \subset I^1_{q+1}} \prod_{i\in I_{q}} \delta^i$. Terms in $R_{I^1_{q+1}}[\partial^{(p,q,0)}_1]$ contain either $\delta^1$, $\delta^2$, or both. Due to the symmetry between $\delta^1$ and $\delta^2$, we have $R_{I^1_{q+1}}[\partial^{(p,q,0)}_1] = (\delta^1 + \delta^2)(\prod_{i\in (I^1_{q+1}/\{1,2\})} \delta^i)$. Therefore, $R_{I^1_{q+1}}[\partial^{(p,q,0)}_1] \partial^{1}_2$ is determined by $\partial^{1}_2 (\delta^1 + \delta^2) = (\delta^1 + \delta^2) (\delta^1 + \delta^2) = 0$. Recognizing that $\delta^1 + \delta^2$ behaves as the boundary operator of a 2D toric code (or more generally, a 2D HGP code), we conclude that each branch of the fork complex effectively embeds a hidden 2D toric code structure, which ensures the satisfaction of the bootstrap equations. This observation is reminiscent of the foliated fracton order theory and the general hierarchy of long-range entanglement patterns~\cite{Shirley2018, Li2023}, although a mathematically rigorous relation remains to be established.

\textit{Outlook.}
In this Letter, we propose the quantum bootstrap product paradigm that constructs full CSS codes by bootstrapping from a segment of the tensor product chain complex of input classical codes. A series of intriguing questions naturally follows. First, while this work mainly discusses products of 1D repetition codes, exploring the QBP of more general classical codes is essential to fully uncover the utility of this paradigm. Second, the mathematical properties of the structure formed by the solutions of the bootstrap equation require further exploration, which is crucial for systematically computing the parameters of QBP codes. Moreover, further generalization of QBP codes using techniques such as balanced products may lead to significantly improved code parameters, which is of great practical interest~\cite{Kovalev2012,Fan2016,Hastings2020,Panteleev2020,Breuckmann2020}. Finally, the correspondence between fork complexes and TD fracton codes suggests that further investigation of QBP codes may shed light on the theoretical understanding of fracton orders and other novel types of topological orders.

\textit{Acknowledgements.}
We thank for the beneficial discussion with Hui Zhai, Yue Wu, Chengshu Li, Yifei Wang, Yu-An Chen and Hao Song. This work is supported by the Shanghai Committee of Science and Technology (Grant No.~25LZ2600800) and  M.-Y. Li is supported by the Shuimu Tsinghua Scholar Program.

\textit{Code availability.}
The codes for this work are available
at \url{https://github.com/yuziyuon/QuantumBootstrapProduct}.

\begin{widetext}
	\centering
	\section*{End Matter}
\end{widetext}

\begin{figure}[t]
	\centering
	\includegraphics[width=\columnwidth]{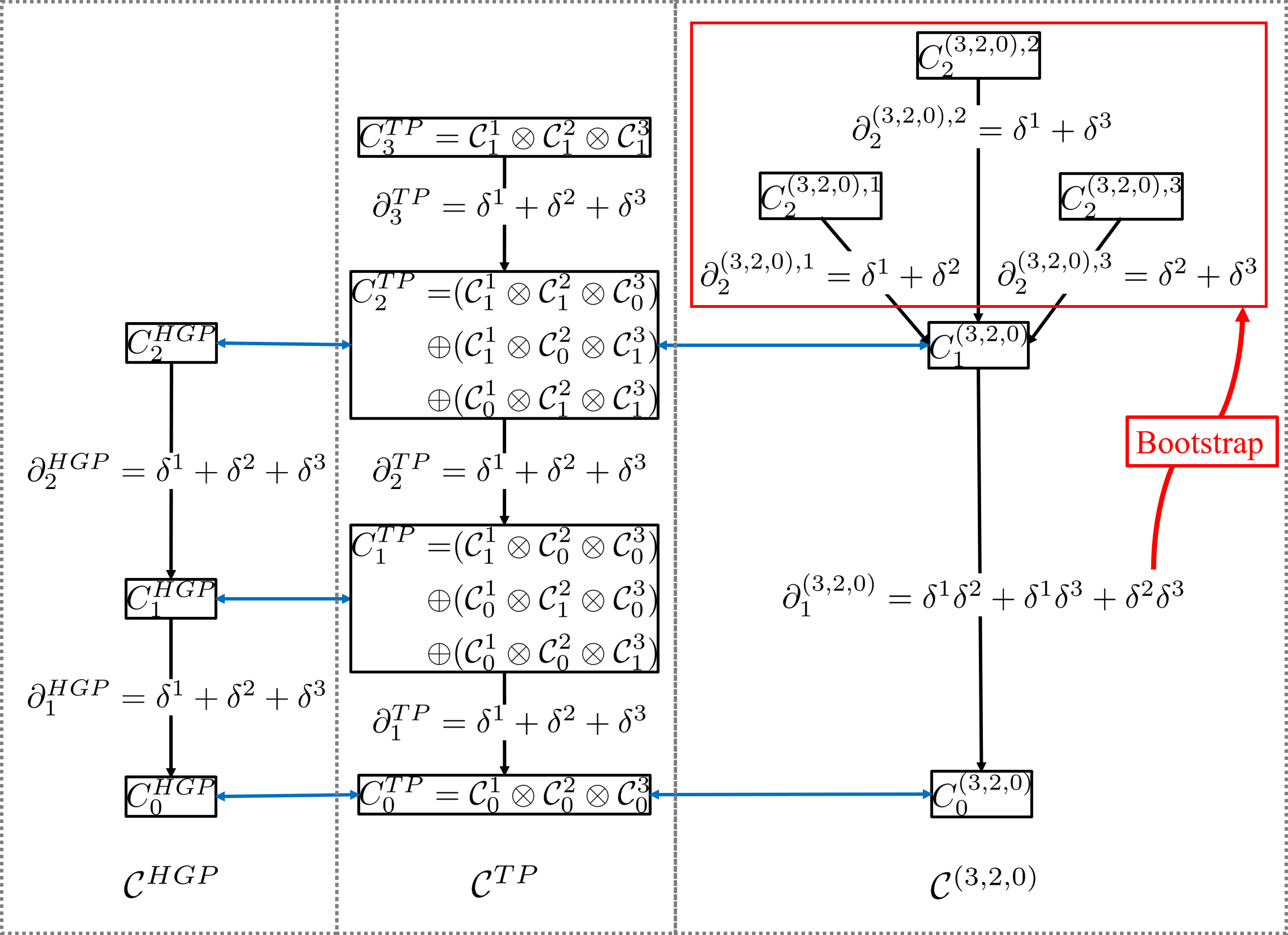}
	\caption{A comparison between the 3D HGP and the $(3,2,0)$ QBP of input classical codes $\C^i: C^i_1\xrightarrow{\delta^i} C^i_0,\ i=1,2,3$. As shown, although both products are based on the tensor product of input classical codes $\C^{TP}$, the HGP formalism extracts a segment of three terms from $\C^{TP}$ to represent the resulting code, whereas the QBP formalism generates the complete resulting code by solving a bootstrap equation. The multiplicity of solutions leads to a fork structure in the resulting complex.}
	\label{fig:products}
\end{figure}

We provide a brief review of the chain complex representation of classical linear and quantum CSS codes, and then use the X-cube code as a concrete example to illustrate the fork complex. An intuitive comparison is illustrated in Fig.~\ref{fig:products}.

\emph{Chain complex representation of CSS codes.}
Classical linear codes and quantum CSS codes are represented by two-term and three-term chain complexes, respectively.

A classical linear code is defined by bits $\{b_i\}$ and parity checks $\{c_j\}$. It corresponds to a chain complex $C_1 \xrightarrow{\delta} C_0$, where $C_1$ and $C_0$ are vector spaces over $\mathbb{F}_2$ spanned by bits and checks, respectively. The map $\delta$ is defined such that $\delta b_i = \sum_{j} c_j$, which sums over all checks involving $b_i$. The code subspace is the kernel $\ker \delta$.

Analogously, a quantum CSS code is represented by $C_2 \xrightarrow{\partial_2} C_1 \xrightarrow{\partial_1} C_0$, where $C_2$, $C_1$, and $C_0$ are vector spaces over $\mathbb{F}_2$ spanned by $Z$-checks, qubits, and $X$-checks, respectively. Here, $\partial_2$ maps a $Z$-check to the sum of qubits involved in it, and $\partial_1$ maps a qubit to the sum of the $X$-checks involving it. The condition $\partial_1 \partial_2 = 0$ guarantees that all $X$- and $Z$-checks commute. Thus, any such complex defines a valid CSS code.

\emph{Fork complex representation of X-cube code.}
We illustrate the fork complex using the X-cube code as an example. It can be constructed as a $(3,2,0)$ QBP of three identical $\C^{1D}$ codes. The resulting complex is denoted by $\C^{(3,2,0)}$.

Using the notation established in the main text, the qubits and $X$-checks span $C^{(3,2,0)}_1 = C^{TP}_2$ and $C^{(3,2,0)}_0 = C^{TP}_0$, respectively. Since $\C^{TP}$ represents a 3D cubic lattice, this relation places qubits on plaquettes ($2$-cells) and $X$-checks on vertices ($0$-cells). The boundary map $\partial^{(3,2,0)}_1 = \delta^1 \delta^2 + \delta^1 \delta^3 + \delta^2 \delta^3$ implies that an $X$-check associated with a vertex involves qubits on the 12 surrounding plaquettes.

Solving the bootstrap equation $R_{I^l_t}[\partial_1] \partial^l_2 = 0$ yields solutions generated by operators of the form $\delta^i + \delta^j$. Specifically, we have three candidate solutions $\partial^{1}_2 = \delta^1 + \delta^2$, $\partial^{2}_2 = \delta^1 + \delta^3$, and $\partial^{3}_2 = \delta^2 + \delta^3$. Due to the linear dependence $(\delta^1 + \delta^2) + (\delta^2 + \delta^3) = \delta^1 + \delta^3$, only two of these are independent. Each branch defines a $Z$-check on every element of $C^{TP}_3$ (cube). For instance, $\partial^{1}_2$ maps a cube to the sum of the qubits on the four plaquettes normal to the first two axes. This means that a basis element of the corresponding subgroup of $C^{(3,2,0)}_2$ represents a $Z$-check defined on a cube that is the product of Pauli $Z$ operators on these four plaquettes. This structure exactly reproduces the X-cube code on the dual lattice.

\clearpage

\onecolumngrid

\setcounter{section}{0}  
\setcounter{figure}{0}   
\setcounter{table}{0}    
\setcounter{equation}{0} 

\renewcommand{\thesection}{S\arabic{section}}  
\renewcommand{\thetable}{S\arabic{table}}  
\renewcommand{\thefigure}{S\arabic{figure}} 
\renewcommand{\theequation}{S\arabic{equation}}

\begin{center}
	\Large\textbf{Supplementary Materials} \\
	\vspace{0.5cm}
\end{center}

\tableofcontents

\section{A brief review of tetra-digit (TD) codes}
\label{sec:td}

Tetra-digit (TD) codes constitute a family of quantum CSS codes defined on (hyper)cubic lattices. Each code is labeled by a quadruple $[d_n,d_s,d_l,D]$, where $d_n<d_s<d_l\leq D$ are four non-negative integers specifying the assignment of qubits and checks~\cite{Li2020,Li2021,Hu2025}. This family encompasses both the paradigmatic topological toric code and the X-cube fracton code, generalizing them to arbitrary dimensions within a unified framework. It thus provides a versatile platform for exploring both topological and fracton codes.

Specifically, a $[d_n,d_s,d_l,D]$ TD code is defined on a $D$-dimensional hypercubic lattice, where qubits and checks are arranged as follows:
\begin{itemize}
	\item Each $d_s$-cell hosts a qubit;
	\item Each $D$-cell is associated with an $X$-check that acts on all neighboring qubits (qubits on the $d_s$-cells incident to the $D$-cell);
	\item Each $d_n$-cell is associated with a set of $Z$-checks; each $Z$-check acts on qubits that are both incident to the $d_n$-cell and fully embedded within the same $d_l$-dimensional subsystem containing the $d_n$-cell.
\end{itemize}

When the four integers satisfy the condition $\binom{d_l-d_n}{d_s-d_n} = 0 \pmod 2$, the defined $X$- and $Z$-checks commute, forming a valid CSS code~\cite{Li2020}. In this work, we focus on cases where $d_l - d_s = d_s - d_n =1$. Under this condition, if $D=d_l$, the definition coincides with a $D$-dimensional toric code with point-like syndromes (note that for $D\geq 4$, multiple types of toric codes exist, and the $[D-2,D-1,D,D]$ codes correspond to a specific type).

\begin{figure}[t]
	\centering
	\includegraphics[width=0.6\columnwidth]{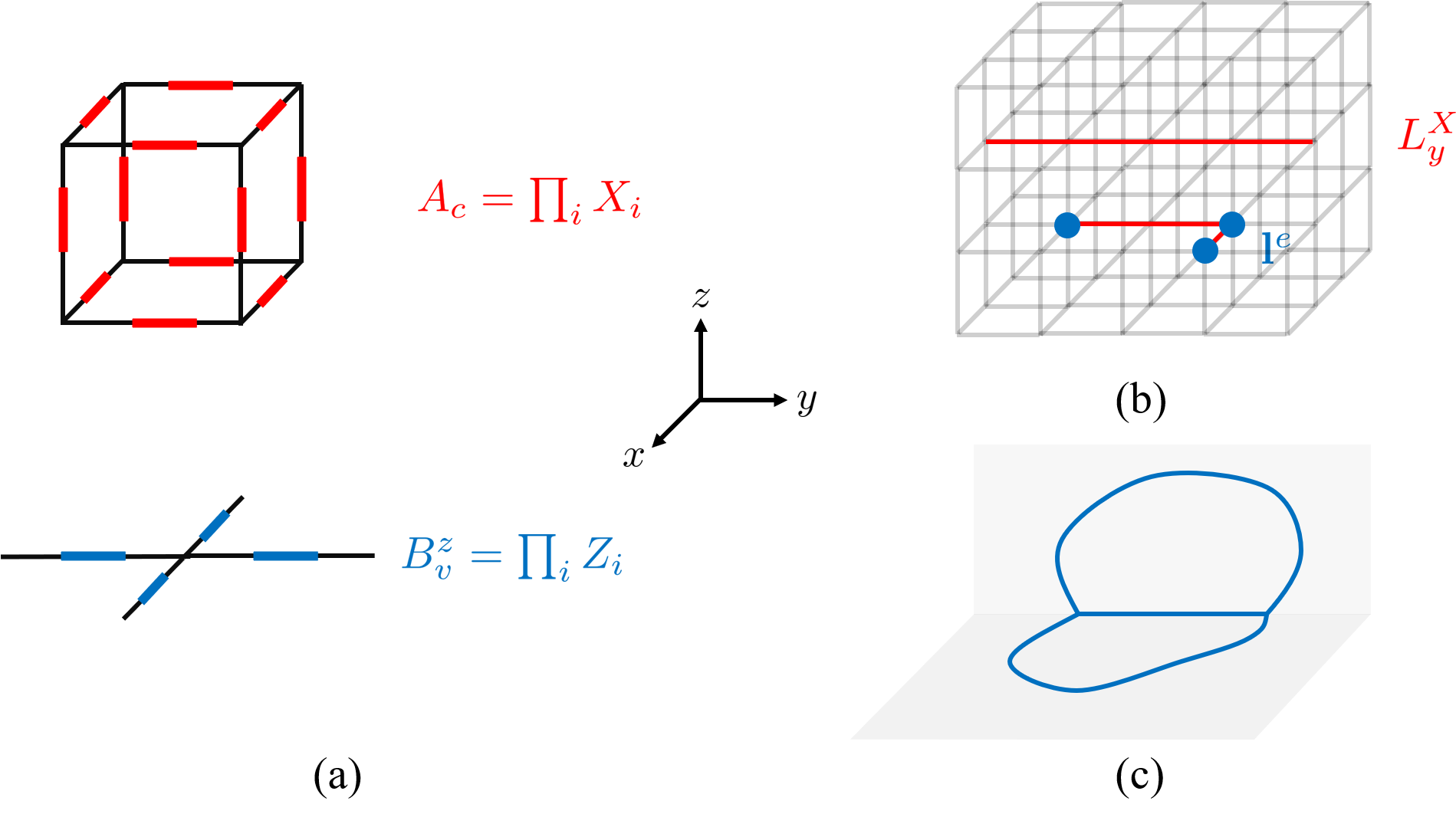}
	\caption{Characteristics of tetra-digit (TD) codes. (a) Illustration of the $X$- and $Z$-checks in the X-cube code, defined on a cube $c$ and a vertex $v$, respectively. Qubits located on links involved in these checks are colored red and blue, respectively. The superscript $z$ of $B^z_v$ denotes a check involving qubits fully embedded in a plane perpendicular to the $z$-direction. (b) shows a noncontractible string logical operator $L^X_y$ composed of Pauli $X$ operators, and three $Z$-syndromes $\mathbf{l}^e$ generated by a string of Pauli $X$ errors. (c) Demonstration of a non-manifold-like syndrome in the $[1,2,3,4]$ code, composed of $Z$-syndromes distributed along the blue strings.}
	\label{fig:td}
\end{figure}

We consider the X-cube code ($[0,1,2,3]$) as a concrete example, representing the simplest nontrivial case where $D>d_l$. This code is defined on a 3D cubic lattice: each link hosts a qubit, and each cube is associated with an $X$-check. Each vertex is associated with three $Z$-checks, where each $Z$-check involves the four qubits on the links that both emanate from the vertex and lie within the same 2D plane (see Fig.~\ref{fig:td} (a)). Note that although the three $Z$-checks at a vertex are not linearly independent (their product is the identity, making one redundant), we retain the full set to preserve the local isomorphism.

Unlike pure topological codes, the syndrome distribution in fracton codes exhibits a mixed dependence on both the topology and the geometry of the error configuration. For instance, a string of Pauli $X$ errors flips $Z$-checks not only at the endpoints of the string but also at its turning points (corners). Such additional syndrome information potentially leads to higher error thresholds, as verified numerically in Refs.~\cite{Song2022,Canossa2025}. Furthermore, although logical operators in the X-cube code are supported on topologically noncontractible strings, their action also depends on their spatial location. Consequently, the code dimension of the X-cube code grows with the system size (see Fig.~\ref{fig:td} (b)). These two features are characteristic of generic fracton codes.

Higher-dimensional TD codes present even more nontrivial features. For example, Pauli $X$ errors distributed in a membrane in the $[1,2,3,4]$ code can create a syndrome with a non-manifold-like shape, such as the structure shown in Fig.~\ref{fig:td} (c). Moreover, since $[0,1,2,D]$ codes possess code dimensions scaling as $k \propto L^{D-2}$, their rate advantage over pure topological codes becomes increasingly pronounced as the spatial dimension $D$ increases.

\section{Polynomial representation of boundary operators}
\label{sec:poly}

In this section, we provide the formal definition of the polynomial representation of boundary operators utilized in the main text. By extending the boundary operators to endomorphisms acting on vector spaces spanned by both bits and checks, we demonstrate that the polynomials correspond to well-defined linear operators on the tensor product space.

\subsection{Extended boundary operators}

To treat boundary maps as algebraic objects that can be multiplied and added, we must define them on a common vector space. For a single classical code $\mathcal{C}^i$, the chain complex consists of two spaces: the space of bits $C^i_1$ and the space of checks $C^i_0$. We define the total local space as the direct sum $V^i = C^i_1 \oplus C^i_0$.

The boundary map $\delta^i$ naturally maps $C^i_1 \to C^i_0$. We extend this map to a linear operator acting on the total space $V^i$ by defining its action on the basis elements:
\begin{itemize}
	\item If $c \in C^i_1$ (a bit vector), then $\delta^i c$ is the standard boundary, i.e., a linear combination of checks involving $c$.
	\item If $c \in C^i_0$ (a check vector), then $\delta^i c = 0$.
\end{itemize}
This extended operator is strictly linear. In a matrix representation where the basis of $V^i$ is ordered as (bits, checks), $\delta^i$ takes the block form:
\begin{align}
	\delta^i_{ext} = \begin{pmatrix} 0 & 0 \\ M_{\delta^i} & 0 \end{pmatrix},
\end{align}
where $M_{\delta^i}$ is matrix representation of $\delta^i$. The zero block in the bottom-right corner ensures that the boundaries of checks vanish. Since this defines a matrix operation, linearity is automatically satisfied.

\subsection{Monomials in the tensor product space}

We now consider a system of $p$ classical input codes. The total global space is the tensor product of the total local spaces: $\mathbf{V} = V^1 \otimes V^2 \otimes \dots \otimes V^p$.

A \textit{monomial} in our polynomial representation, denoted by $\delta^S$ for a subset of indices $S \subset I_p$, is defined as the tensor product of operators acting on this global space:
\begin{align}
	\label{s_eq:monomial}
	\delta^S \equiv \bigotimes_{j=1}^p \hat{O}^j,
\end{align}
where the operator $\hat{O}^j$ acting on the $j$-th classical code is given by:
\begin{align}
	\hat{O}^j = \begin{cases} 
		\delta^j & \text{if } j \in S, \\
		\mathbb{I}^j & \text{if } j \notin S.
	\end{cases}
\end{align}
Here, $\mathbb{I}^j$ is the identity operator on $V^j$. 

Crucially, $\delta^S$ is a tensor product of linear maps and is therefore a linear map itself. Its action on a basis tensor $c = c_1 \otimes \dots \otimes c_p$ computes the boundaries of the components specified by the set $S$:
\begin{align}
	\delta^S(c_1 \otimes \dots \otimes c_p) = (\hat{O}^1 c_1) \otimes \dots \otimes (\hat{O}^p c_p).
\end{align}
This definition rigorously justifies the case-by-case behavior described in the main text. For instance, if $j \in S$ but the input component $c_j$ is a check vector ($c_j \in C^j_0$), the entire term vanishes because the linear operator $\delta^j$ maps checks to the zero vector.

\subsection{Polynomials as sums of linear maps}

Finally, we demonstrate the validity of polynomial representations of boundary operators in the QBP paradigm. As an example, we consider $\partial^{(p,q,w)}_1$, which is expressed as a sum of monomials:
\begin{align}
	\partial^{(p,q,w)}_1 = \sum_{S \subset I_p, |S|=q-w} \delta^S.
\end{align}
Since each $\delta^S$ is a linear map on $\mathbf{V}$, their sum $\partial^{(p,q,w)}_1$ is also a linear map. Although $\partial^{(p,q,w)}_1$ is defined on the global space $\mathbf{V}$, we are specifically interested in its action on the \textit{qubit subspace} $C^{(p,q,w)}_1$. This subspace is a direct sum of specific tensor product sectors (those containing exactly $q$ bit components). When the polynomial acts on a vector in $C^{(p,q,w)}_1$, linearity ensures it distributes over the vector sum. For any valid basis vector in the qubit space, only the terms in $\partial^{(p,q,w)}_1$ that are compatible with the support of the basis vector yield non-zero results, correctly mapping the basis vector into the $X$-check space $C^{(p,q,w)}_0$. The analysis of the $\partial^l_2$ boundary operators follows similarly.

\section{A general algorithm for solving bootstrap equations}
\label{sec:algorithm}

In this section, we describe an algorithm used to solve bootstrap equations. Building upon the polynomial representation, the problem of finding valid boundary operator components $\partial^l_2$ is reduced to finding polynomials in the variables $\{\delta^i\}$ that annihilate the restriction of $\partial^{(p,q,w)}_1$ on a support $I^l_t$.

We seek a complete set of solutions $\partial^l_2$ satisfying the bootstrap equation:
\begin{align}
	R_{I^l_t}[\partial^{(p,q,w)}_1] \partial^{l}_2 = 0,
\end{align}
where $I^l_t \subset I_p$ is a support set of size $t$. Our algorithm proceeds iteratively by increasing the support size $t$ from $q+1$ to $p$. At each level $t$, we perform three main steps: first, identify the solution space; second, remove redundant solutions; finally, optimize the weight (i.e., the number of monomial terms) of the solutions.

\subsection{Solving the bootstrap equation for a given support set}
For a fixed support size $t$, we consider a subset of indices $I_t \subset I_p$. Without loss of generality, we focus on the subset $I_t = \{1, 2, \dots, t\}$. The restriction of the boundary operator $\partial^{(p,q,w)}_1$, $\tau \equiv R_{I_t}[\partial^{(p,q,w)}_1]$, retains only those terms in $\partial^{(p,q,w)}_1$ composed entirely of variables in $I_t$. $\tau$ corresponds to the elementary symmetric polynomial of degree $k = q-w$ on the variables $\{\delta^i\}_{i \in I_t}$:
\begin{align}
	\tau = \sum_{S \subseteq I_t, |S|=k} \delta^S,
\end{align}
where $\delta^S$ follows the definition in Eq.~\ref{s_eq:monomial}.

Let $\mathcal{B}_{d_{sol}}$ be the basis of all valid monomials of degree $d_{sol}= t-q$ on $I_t$, which is of the size $\binom{t}{d_{sol}}$. Representing a polynomial $\xi$ as a vector in this basis, the multiplication by $\tau$ acts as a linear map from the space of degree-$d_{sol}$ polynomials to the space of degree-$(d_{sol}+k)$ polynomials. We denote the matrix representation of this map by $M_{\tau}$. The condition $\tau \xi = 0$ corresponds to the linear equation over $\mathbb{F}_2$:
\begin{align}
	M_{\tau} \xi = 0.
\end{align}
The solution space $V_{sol}(I_t)$ corresponds to the null space of $M_{\tau}$, which can be obtained by using Gaussian elimination over $\mathbb{F}_2$.

\subsection{Filtering primitive generators}
Simply solving the bootstrap equation leads to an overcomplete set of solutions. For efficiency, we require only the \textit{primitive generators} of the solution spaces at each level: namely, solutions that are intrinsic to the support size $t$, rather than trivial extensions of solutions from smaller support sizes $t' < t$.

Let $\mathcal{G}_{t'}$ be the set of primitive generators identified at a smaller size $t'$. A generator $g \in \mathcal{G}_{t'}$ defined on $I_{t'} \subset I_t$ can be embedded into the larger support $I_t$. Such an embedding can be realized by multiplying a monomial $m$ composed of the new variables in $I_t \setminus I_{t'}$, such that the total degree matches $t-q$. The subspace of redundant solutions, $V_{red}(I_t)$, is spanned by all such products:
\begin{align}
	V_{red}(I_t) = \text{span} \left\{ g \cdot m \;\middle|\; g \in \bigcup_{t'<t} \mathcal{G}_{t'}, \deg(g \cdot m) = t-q \right\}.
\end{align}

We define the space of primitive generators at level $t$ as the quotient space:
\begin{align}
	V_{prim}(I_t) = V_{sol}(I_t) / V_{red}(I_t).
\end{align}
Computationally, this is achieved by forming a matrix whose rows include basis vectors of $V_{red}(I_t)$ followed by basis vectors of $V_{sol}(I_t)$. By performing row reduction, we identify the vectors in $V_{sol}(I_t)$ that are linearly independent of $V_{red}(I_t)$. These vectors form the raw basis for the primitive generators.

\subsection{Optimizing primitive solutions}
The basis for $V_{prim}(I_t)$ obtained from linear algebra is not unique. To simplify the resulting quantum code, we apply a greedy algorithm to lower the weights of the checks.

Let $\mathcal{G} = \{g_1, \dots, g_m\}$ be the basis of polynomials obtained from the previous step. The greedy algorithm iteratively attempts to replace a basis element $g_i$ with the sum $g_i + g_j$ (for some $j \neq i$) if and only if:
\begin{align}
	\text{weight}(g_i + g_j) < \text{weight}(g_i).
\end{align}
This procedure optimizes the complexity of the polynomial representation of $\partial^l_2$, which directly corresponds to reducing the weight (density) of the checks in the resulting quantum code.

The final output of the algorithm is a collection of optimized primitive generators $\partial^l_2$ for all support sizes $t$, which fully specifies the boundary map $\partial^{(p,q,w)}_2$ of the QBP code.

\end{document}